# *Low threshold anti-Stokes Raman laser on-chip*

HYUNGWOO CHOI[1], DONGYU CHEN[2], FAN DU[1], RENE ZETO[1], ANDREA ARMANI[1,2,*]

[1]*Mork Family Department of Chemical Engineering and Materials Science, University of Southern California, Los Angeles, California 90089, USA*
[2]*Ming Hsieh Department of Electrical Engineering-Electrophysics, University of Southern California, Los Angeles, California 90089, USA*
*\*Corresponding author: armani@usc.edu*

**Raman lasers based on integrated silica whispering gallery mode resonant cavities have enabled numerous applications from telecommunications to biodetection. To overcome the intrinsically low Raman gain value of silica, these devices leverage their ultra-high quality factors (Q), allowing sub-mW stimulated Raman scattering (SRS) lasing thresholds to be achieved. A closely related nonlinear behavior to SRS is stimulated anti-Stokes Raman scattering (SARS). This nonlinear optical process combines the pump photon with the SRS photon to generate an upconverted photon. Therefore, in order to achieve SARS, the efficiency of the SRS process must be high. As a result, achieving SARS in on-chip resonant cavities has been challenging due to the low lasing efficiencies of these devices. In the present work, metal-doped ultra-high Q (Q>10[7]) silica microcavity arrays are fabricated on-chip. The metal-dopant plays multiple roles in improving the device performance. It increases the Raman gain of the cavity material, and it decreases the optical mode area, thus increasing the circulating intensity. As a result, these devices have SRS lasing efficiencies that are over 10x larger than conventional silica microcavities while maintaining low lasing thresholds. This combination enables SARS to be generated with sub-mW input powers and significantly improved anti-Stokes Raman lasing efficiency.**

## 1. INTRODUCTION

Over the past decade, lasers have enabled numerous discoveries in fundamental science and have played a key role in many technological advances. These systems have relied on synergistic innovations in optical devices and in nonlinear optical materials. For example, high harmonic femtosecond pulsed laser systems have unraveled the kinetics of chemical reactions [1–3], and Raman-based fiber lasers form the foundation for a wide range of medical [4–6] and telecommunication systems [7–9]. However, there is an ever increasing demand to move from benchtop to on-chip platforms.

To meet this demand, recent research efforts have investigated creating the analog of the Raman fiber lasers using on-chip whispering gallery mode optical resonators as the lasing cavity [10–14]. Raman lasers rely on the third order nonlinear optical processes that result from the interaction of light with the vibrational modes, or optical phonons, of the scattering medium [15,16]. Because optical resonators can have quality factors (Q) in excess of 100 million, they can support large build-up intensities and achieve low Raman lasing thresholds [10,16]. However, because the Raman gain of the device material is low, the stimulated Raman scattering (SRS) lasing efficiency is poor. This low efficiency makes it extremely challenging to achieve a related nonlinear effect, simulated anti-Stokes Raman scattering (SARS), in an integrated device platform [17].

Similar to SRS, SARS is also a third order nonlinear effect. However, SARS is dependent on the interaction between the pump photons and the SRS signal, as shown in Fig. 1 (a) [18]. Thus, to generate SARS, there must a sufficient population of photons both at the SRS wavelength and the pump wavelength. Therefore, in integrated devices, the intensity of the SARS emission is usually several orders of magnitude weaker than that of SRS [19,20].

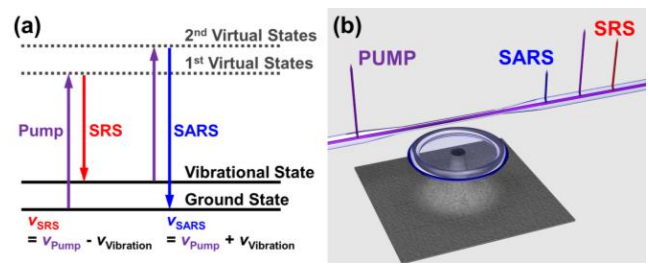

**Fig. 1.** (a) Energy level diagram of red-shifted stimulated Raman scattering (SRS) and blue-shifted stimulated anti-Stokes Raman scattering (SARS) with the vibrational state of the gain medium. $\nu_{Pump}$, $\nu_{SRS}$, $\nu_{SARS}$, and $\nu_{Vibration}$ are corresponding to the frequencies of pump, SRS, SARS, and vibrational optical phonon, respectively. (b) Schematic image of generated SRS (red) and SARS (blue) with Pump (purple) in on-chip metal-doped silica hybrid toroid resonator with tapered optical fiber waveguide.

In previous works, one successful approach to generate SARS is to use a very high power laser, increasing the number of photons available for SARS [21–24]. Additionally, efforts developing nanomaterials that can generate Anti-Stokes emissions have been rapidly progressing [25]. However, both methods will face hurdles to being readily integrated on-chip. Recent research using metal-doped silica toroidal whispering gallery mode microcavity-based microlasers has shown a possible alternative strategy (Fig. 1 (b)). Metal dopants enable silica sol-gel layers to exhibit high Raman gain, enhancing Raman lasing efficiencies as high as 40% with lasing thresholds in the sub-mW range [26,27]. These devices could provide a path to overcome the previous hurdles and be the ideal on-chip platform for generating SARS with single pump source.

In this work, we demonstrate the generation of SRS and SARS using Zr-doped and Ti-doped silica microresonators fabricated on silicon. The experimental results clearly indicate that the metal-doped silica hybrid devices show significant enhancement in both SRS and SARS efficiencies with reduced thresholds when compared to the undoped devices. The theoretical analysis of SARS behavior in the devices is developed by combining analytical theory with finite element method simulations, and the results reveal that the ability to generate SARS with sub-mW thresholds can be attributed to improved mode confinement and increased Raman gain caused by the metal dopants.

## 2. THEORETICAL ANALYSIS

### A. Theoretical motivation

Previous work has shown that the anti-stokes wave is generated through a four-wave mixing process involving the pump mode and the Raman mode [28]. As the amplitude of the Raman mode is several orders of magnitude higher than the anti-stokes mode [19,20], the coupled mode equations for the amplitude of the anti-Stokes mode ($A_A$) can be written as:

$$A_A = \frac{2\gamma}{\Delta\omega} A_P A_P A_R^* \left(e^{i\Delta\omega t} - 1\right) \quad (1)$$

, where $\gamma = (\omega_P\, n_2)/(c\, A_{eff})$, $\omega_P$ is the pump frequency, $n_2 \approx 2.2 \times 10^{-20}\, m^2/W$ is the Kerr nonlinear coefficient for silica, c is the speed of light in vacuum, and $A_{eff}$ is the effective mode area. $A_P$ and $A_R$ are the amplitude of the pump and the stimulated Raman mode, respectively. $\Delta\omega = 2\omega_P - \omega_R - \omega_A$ is the frequency detuning, where $\omega_R$ and $\omega_A$ are the frequency of the Raman and the anti-Stokes modes, respectively. $t$ is the interaction time between the pump and the Raman modes. The intensity of electric field (I) is related to the amplitude of the field (A) according to: I = ε $A^2$/ 2 where ε is the permittivity of the medium. By solving the Equation (1) with an initial condition ($t$ = 0, $A_A$ = 0), the intensity of the anti-Stokes ($I_{SARS}$) can be described as:

$$I_{SARS} = \left(\frac{4\, w_p\, n_2\, t}{\varepsilon\, c}\right)^2 \left(\frac{\sin(\Delta\omega\, t/2)}{(\Delta\omega\, t/2)}\right)^2 I_{Pump}^2 \frac{I_{SRS}}{A_{eff}^2} \quad (2)$$

Here, $I_{Pump}$ and $I_{SRS}$ are the intensity of the pump and the Raman modes, respectively. Above the Raman threshold, a clamped pump field is observed. As the pump mode is a clamped mode, $I_{Pump}$ can be taken as a constant and is independent of the coupled power [28]. This assumption modifies the typical relationship between the coupled power into the resonator and the launched fiber power, which gives a square root dependence of the pump-to-Raman conversion. In other words, due to the clamped mode, the coupled power is constant even when the amount of power in the fiber increases.

Based on Equation 2, there are several relationships that are important to highlight. First, $I_{SRS}$ is linearly dependent on the coupled power, as has been shown in previous work with resonant cavity devices. Second, because $I_{SARS}$ is linearly proportional to $I_{SRS}$, the intensity of the anti-stokes wave is also linearly dependent on the coupled power.

Lastly, it worthwhile investigating the dependence of $I_{SARS}$ on the optical mode area. Typically, mode area is overlooked when evaluating whispering gallery mode device types, as quality factors can vary by several orders of magnitude and thus have a larger impact on the circulating intensity. However, in the present work, the optical quality factor is held constant, but the optical mode area is a variable. From equation 2, it is clear that $I_{SARS}$ is inversely dependent on the $A_{eff}^2$. In other words, a smaller mode area results in a higher intensity of SARS emission. In contrast, the $A_{eff}$ has a linear relationship with the SRS threshold and is inversely proportional to the intensity of SRS. To calculate mode area, finite element method analysis is needed.

### B. COMSOL Multiphysics FEM simulations

Given the dependence of SRS and SARS on $A_{eff}$, it is judicious to model the $A_{eff}$ of different device types before performing experiments. This modeling was performed using COMSOL Multiphysics finite element method (FEM) **[29]**. Fig. 2 (a) contains the cross-sectional images of toroids with the fundamental optical modes of both undoped and metal-doped coated silica hybrid resonators. All values used in the model are based on the devices used in the experimental work. Due to the high index of the metal-doped coating layer, the optical mode shifts towards the edge of the toroid and is more tightly confined in the coating layer, causing a decrease in the $A_{eff}$ **[30]**.

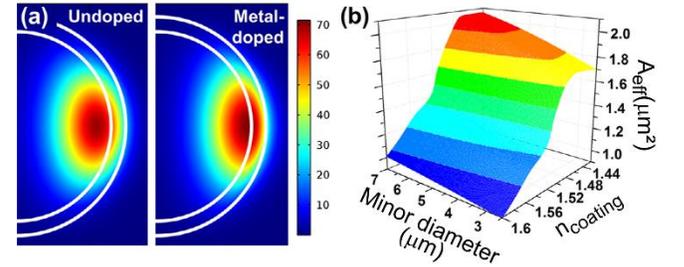

**Fig. 2.** COMSOL Multiphysics finite element method (FEM) simulation results. (a) Optical field distribution in the coated toroidal microcavity. The major and minor diameters of the silica toroid are 55 μm and 6 μm, respectively. The refractive indices of coatings are 1.454 for undoped and 1.520 for metal-doped layer (either Zr or Ti) with thickness of 400 nm. (b) The dependence of the fundamental mode area as a function of the refractive index of coating ($n_{coating}$) and the minor diameter. The coating thickness is fixed to 400 nm, and the wavelength is 1550nm.

The values of all parameters were varied to evaluate the experimental landscape, and the results are plotted in Fig. 2 (b). It is clear that changes in the coating index have a significant effect while changes in the device geometry have negligible impact on mode area over the values modeled.

These simulations can be used to evaluate the impact of changing the film on $A_{eff}$. For example, as the $n_{coating}$ increases from 1.454 to 1.520 at the minor diameter of 6 μm, the $A_{eff}$ decreases from 1.17 to 0.90 μm$^2$. This change would result in an increase of SARS Intensity of 1.69 times or 169%. Thus, it is clear that by adding a thin layer of a high index coating, the ability to generate SARS should be significantly increased, even if the Raman gain remained low.

To experimentally verify this hypothesis, metal-doped sol-gel coated toroidal microcavities are fabricated. The devices consist of a silica toroid cavity with a metal-doped silica sol-gel coating layer. Two different metals are studied in the present work: Zirconium (Zr) and Titanium (Ti) as well as control devices.

## 3. EXPERIMENTAL METHODS

### A. Synthesis of metal-doped silica sol-gel

The metal-doped silica sol–gels are synthesized with an acid-catalyzed hydrolysis–condensation reaction [31]. Either Zr or Ti is selected as the metal dopant because their tetrahedral oxygen bonds form a stable matrix with the silica sol-gel. Methyltriethoxysilane (MTES, Sigma-Aldrich, 98%), which is a silica precursor, is added to ethanol (Decon Laboratories, KOPTEC 200 Proof) with 5 minutes of stirring. Hydrochloric acid (HCl, EMD, 38.0%) is added to this solution to initiate the hydrolysis reaction. After 20 minutes, 10 mol% of either Zr propoxide (Sigma-Aldrich, 70 wt % solution in 1-propanol) or Ti butoxide (Sigma-Aldrich, 97%) to silica is added to the solution, and the reaction progresses for another 2 hours. After the hydrolysis is complete, the sol–gel is aged at room temperature for 24 hours and filtered through a 0.45 μm syringe filter. The solution can be stored in a refrigerator (~5 °C) until it is used.

The doping concentrations used here were determined in previous works [26,27] and represent a balance between achieving the maximum increase in Raman gain and in refractive index with the minimum increase in optical loss. An additional consideration is the ability of the sol-gel matrix to support large concentrations of dopants without cracking, which will decrease the Q.

### B. Material characterization

To analyze the material properties and verify dopant incorporation, it is necessary to fabricate a series of control wafers. These are fabricated by spin-coating the sol-gels onto bare silicon wafers using the same method as in the device protocol. The refractive indices and film thicknesses are measured with an ellipsometer. The film thicknesses are constant (~400 nm) in all samples. The refractive indices of the coating ($n_{coating}$) increase from 1.454 for undoped coating to 1.520 for the 10 mol% of either Zr or Ti-doped coating [26,27].

The Raman spectra of undoped, Zr- and Ti-doped sol-gel layer on control wafer are collected via Reinshaw InVia Raman spectrograph with a 532 nm excitation laser. Silica generally has two Raman responses at 500 and 1000 cm$^{-1}$, corresponding to the bending and stretching mode of silica, respectively [32]. The ratio of two peaks indicates the degree of polarization ($I_P$) of silica. The $I_P$ values increase from 2.96 (undoped) to 4.87 (10 mol% metal-doped) [26,27], indicating that silica become more polarized with metal dopants. In silica, both SRS and SARS are based on the vibration of phonon. In other words, the enhancement of $I_P$ value indicates that silica matrix becomes more susceptible to vibrate with incident electric field, enabling stronger Raman responses. Therefore, with the high Raman gain material, metal-doped silica sol-gel layer can improve not only SRS but also SARS efficiency and threshold.

### C. Device fabrication

Bare silica toroidal micro-resonators are fabricated from 2 μm of thermally grown silica on a Si wafer using three main steps: photolithography and BOE etching to define silica circles pattern on silicon wafers, XeF$_2$ etching to remove silicon isotropically to obtain a suspended silica disk on a silicon support pillar, and CO$_2$ laser reflow to produce silica toroids [33]. All devices fabricated have similar major and minor diameters (55 μm and 6 μm).

After fabrication, the bare silica toroids are treated with an oxygen plasma. Then, the metal-doped silica sol–gel solutions are spin-coated at 7000 rpm for 60 seconds directly onto the devices. After the spin coating, the samples are dried at 80 °C for 5 minutes to evaporate the residual ethanol solvent. The final metal-doped silica hybrid devices are obtained after annealing the samples at 1000 °C for 1 hour in a tube furnace in an ambient environment. This process completes the condensation reaction of the metal-doped silica sol-gel. Fig. 3 (a) shows one of the metal-doped silica hybrid devices taken via scanning electron microscope (SEM).

The presence of metal dopants (either Zr or Ti) is verified using SEM-EDS (energy-dispersive X-ray spectroscopy) measurements as shown at Fig. 3 (c) and (d). Notably, these measurements were performed on the devices with either Zr-doped or Ti-doped thin film coatings, not control wafers. In addition to the peaks originating from the silicon (Si at 1.740 keV) and oxygen (O at 0.523 keV) intrinsic to the Si and SiO$_2$, each spectrum contains peaks corresponding to either Zr (at 2.042 keV) or Ti (at 4.510 keV). The relative intensity of each signal indicates the relative concentration of the element. Considering the intensity of the Zr or Ti signals, the concentration of each element is fairly low compared with Si and O, as expected.

### D. Device characterization

Fig. 3 (b) shows a schematic image of the testing set-up for the silica hybrid resonators. The quality factor (Q) of the device is characterized using a Newport velocity tunable laser centered at 1550 nm. Light is coupled into a device using a tapered optical fiber waveguide [33]. The end of the tapered optical fiber is connected to an optical splitter; one output goes to an oscilloscope to measure the cavity Q of the device (10 %), and the other end is attached to the optical spectrum analyzer (OSA) to detect the optical emissions from device (90 %).

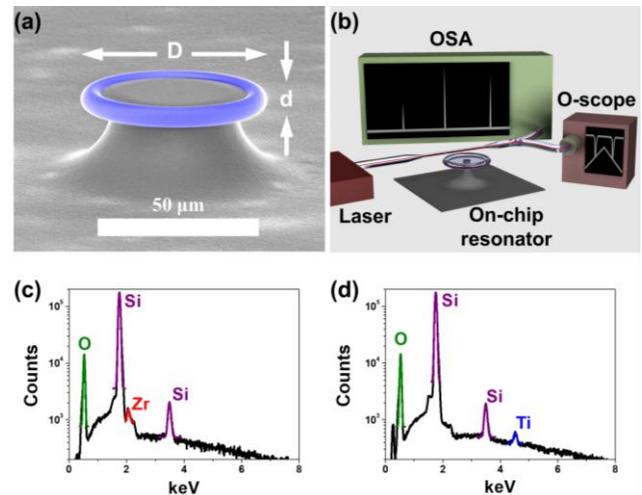

**Fig. 3.** (a) SEM image of a sol-gel coated device indicating major (D) and minor (d) diameters. (b) Schematic image of testing set-up with a laser, a tapered optical fiber, a photodetector (PD) connected to an oscilloscope (O-scope), an optical spectrum analyzer (OSA), and on-chip silica microcavity. EDS spectra from (c) Zr- and (d) Ti-doped silica hybrid devices. The silicon (purple) has peaks at 1.740 and 3.49 keV. The oxygen (green) has a peak at 0.523 keV. The Zr (red) and Ti (blue) have peaks at 2.042 and 4.510 keV, respectively. Each peak is fitted to a Gaussian.

The loaded Q is determined by measuring the transmission spectrum from the oscilloscope and using the expression: $Q = \lambda / \Delta\lambda$, where $\lambda$ is the resonant wavelength of the resonator and $\Delta\lambda$ is the full-width at half maximum (FWHM) of the resonant peak fitted to a Lorentzian [34]. Using a coupled waveguide-cavity model, the intrinsic Q can be determined.

The efficiency and threshold of SRS and SARS behavior is determined by varying the input power and measuring the output SRS and SARS emission on the OSA. The x-intercept of these results represents the threshold, and the slope of the fitted lines is the efficiency of the process. In addition, the nonlinear mechanism giving rise to the observed emission wavelengths can be confirmed by measuring the Stokes and anti-Stokes frequencies and comparing them with the known values of silica.

# 4. RESULTS AND DISCUSSION

## A. Optical Quality Factors (Q)

Fig. 4 contains intrinsic Q values from a series of devices. Intrinsic Q values of the metal-doped devices are slightly lower than those of undoped silica due to the material absorption loss from the metal dopants [30]. Even though the intrinsic Q values of the metal-doped devices are lower than those of the undoped devices, the intrinsic Q values are still above $10^7$. At these device geometries and wavelengths, this Q range represents circulating optical intensities of between ~64.21 GW/cm$^2$ (Q=$10^7$) to ~642.06 GW/cm$^2$ (Q=$10^8$) for the metal-doped devices with 1 mW of input power, which should be more than sufficient to achieve SRS [10,16]. All values in Fig. 4 are summarized in Table 1.

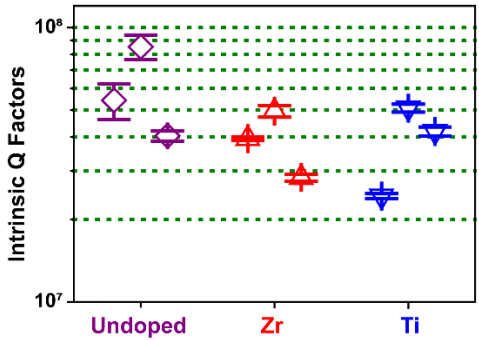

**Fig. 4.** The intrinsic Q of a series of undoped, Zr-doped, and Ti-doped silica hybrid devices.

## B. SARS emission characterization

Figure 5 (a) – (c) show representative spectra for first order SRS and SARS emissions from undoped, Zr-, and Ti-doped devices acquired with a similar coupled power, approximately 1.5 mW, measured via the OSA. The spectra contain the red-shifted SRS peaks and the blue-shifted SARS peaks as well as the pump laser located at the center of the spectra. In the zoom-in spectra, the SARS emission is clearly evident in both metal-doped devices, whereas it is necessary to magnify the y-axis by 10x in order to detect the SARS emission from the undoped device, (Fig. 5 (a) inset). This improvement indicates that the metal-dopants have a significant effect on the SARS generation.

All Stokes and anti-Stokes shifts measured for all devices are summarized in Fig. 6. As expected for a Raman process, the Stokes- and anti-Stokes shift values are the same, and the values range between 12~15 THz for both undoped and metal-doped films [35]. This range falls within the broad, continuous Raman gain spectrum of silica and indicates that the signal in all devices is being generated by the Si-O-Si vibration. The values are included in Table 1.

While the relative magnitudes of the SARS emissions in the spectra in Fig. 5 are indicative of an enhancement, it is necessary to acquire the efficiency and threshold measurements of the SRS and SARS signals to quantify the performance improvement. Figure 7 contains these characterization measurements for the devices plotted in Fig. 5 as a function of coupled power and the results for all devices are summarized in Fig. 8.

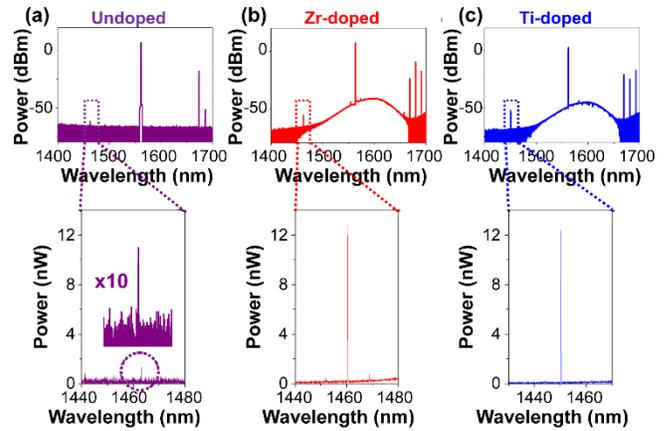

**Fig. 5.** Emission spectra of (a) undoped, (b) Zr-, and (c) Ti-doped devices with similar coupled power (~ 1.5 mW) into the devices obtained via OSA. Zoom-in spectra below show generated SARS.

As can be seen in Fig. 7(a), above threshold, the dependence of SRS output power on input power is linear. These results are in agreement with previous studies on resonant cavity based Raman lasers [16]. In addition, the SARS intensity depends linearly on the coupled power, as expected in Equation (2). Based on these results, the relationship between SRS and SARS emissions can be calculated (Fig. 7(c)). All findings support the predicted linear dependence from Equation (2). In the case of the undoped device, the plotted values in Fig. 7(c) are confined in a small range because of the low efficiency of generating SRS which results in poor SARS generation.

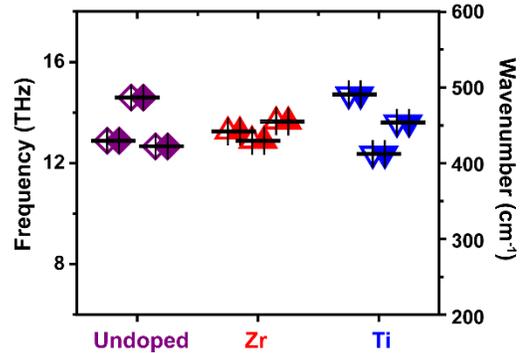

**Fig. 6.** Generated SRS (hollow symbol) and SARS (solid symbol) shift from various devices. As expected, the Stokes and anti-Stokes shifts are identical, providing evidence that the upconverted photons are the result of the anti-Stokes process.

From the data in Fig. 7 and similar data for other devices, the lasing efficiency (slope) and threshold power (x-intercept) are calculated from linearly fitted line. These results are plotted in Fig. 8 and summarized in Table 1. As can be seen, the results are reproduced across multiple devices.

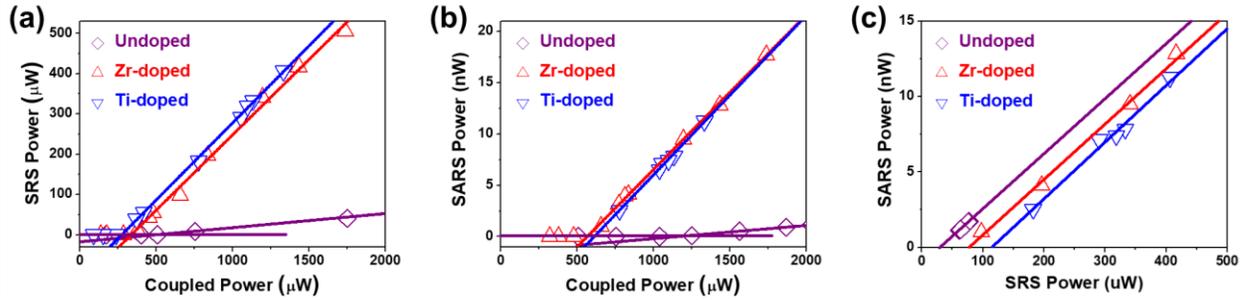

**Fig. 7.** (a) SRS and (b) SARS power as a function of the coupled power into the devices. (c) The ratio of SARS versus SRS power from various devices.

The average SRS threshold values are 573.67 ± 54.35 µW (undoped), 329.33 ± 24.95 µW (Zr-doped), and 342.67 ± 56.98 µW (Ti-doped). As shown in Fig. 8 (b), the average SRS efficiencies are 3.38 ± 0.17% (undoped), 36.22 ± 2.44% (Zr-doped), and 33.22 ± 4.72% (Ti-doped) devices. Thus, the presence of the metal-doped silica sol-gel coating decreases the threshold by over 1.5x and increases the efficiency by 10x. Similar performance improvements due to the inclusion of metal dopants have been observed in previous work and has been attributed to an increase in the polarizability of the silica, which increases the Raman gain, as well as an increase in the refractive index of the metal-doped layer [26,27].

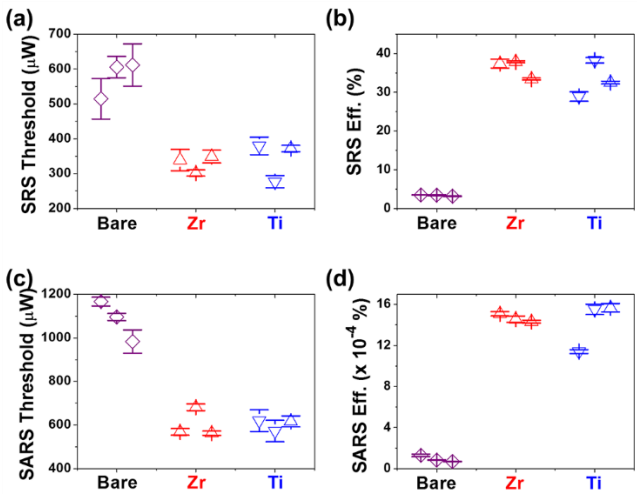

**Fig. 8.** (a) Threshold and (b) efficiency values of SRS, and (c) threshold and (d) efficiency values of SARS from various devices.

This improvement in SRS generation efficiency and threshold enables SARS to be easily generated at low input powers. As seen in Fig. 8 (c), the SARS threshold decrease parallels the decrease observed for SRS, with average SARS threshold values of 1,081.49 ± 92.51 µW (undoped), 604.37 ± 66.83 µW (Zr-doped), and 603.12 ± 26.75 µW (Ti-doped). This threshold improvement is ~1.8, which is comparable to the improvement in the SRS generation.

The average SARS efficiencies also increased with the addition of the coating layer and are 0.94 ± 0.31, 14.66 ± 0.40, and 14.20 ± 2.42 x $10^{-2}$ % for undoped, Zr-, and Ti-doped devices, respectively. Therefore, the improvement due to the addition of the metal-doped thin films is ~15x. This efficiency improvement is due to the increase in the population of the vibrational state in the metal-doped devices (Fig. 1 (a)), allowing more pump photons to be excited from the vibrational state to the 2nd virtual state.

It is worth noting that this 15x efficiency improvement is larger than the 10x SRS efficiency increase. The larger effect can be directly attributed to a decrease in $A_{eff}$ in Equation (2). With all other factors in Equation (2) constant, the larger SARS efficiency enhancement is due to the multiplicative effect of the SRS efficiency increasing and the effective mode area decreasing.

## 5. CONCLUSIONS

In conclusion, low threshold anti-Stokes Raman lasers are demonstrated using an integrated resonant cavity platform. To achieve this on-chip platform, simultaneous innovation in device design and in material performance were achieved. By coating an optical cavity with a metal-doped thin film, the material Raman gain was increased. Additionally, the material developed was designed to have a low optical loss, thus allowing the ultra-high-Q factor to be maintained, and the device geometry and material index were optimized to reduce the optical mode area. As a result of these synergistic improvements, the efficiency of the generated SARS is improved over 15x and the threshold is decreased over 1.8x as compared to other on-chip silica devices. On-chip SARS generators will find numerous applications such as in the development of on-chip Raman systems biochemical detection [36–38] and unraveling quantum mechanical behavior [39,40].

**Funding Information.** The work was supported by Northrop Grumman and the Office of Naval Research (N00014-17-1-2270).

**Acknowledgment**. We would like to thank Andre Kovach for valuable discussions on data analysis.

**Table 1.** Summary of all results

|  |  | Bare | Bare | Bare | Zr | Zr | Zr | Ti | Ti | Ti |
|---|---|---|---|---|---|---|---|---|---|---|
| Diameter (µm) | | 53.42 | 55.20 | 55.73 | 53.53 | 53.80 | 51.60 | 52.67 | 52.07 | 53.80 |
| Intrinsic Q (x $10^7$) | | 5.43 ± 0.80 | 8.52 ± 0.88 | 4.03 ± 0.17 | 3.94 ± 0.06 | 4.96 ± 0.24 | 2.84 ± 0.08 | 2.43 ± 0.05 | 5.09 ± 0.17 | 4.18 ± 0.16 |
| Shift of SARS & SRS | cm$^{-1}$ | 429.63 | 462.32 | 422.29 | 430.30 | 483.00 | 454.65 | 493.34 | 411.95 | 453.65 |
| | THz | 12.88 | 13.86 | 12.66 | 12.90 | 14.48 | 13.63 | 14.79 | 12.35 | 13.60 |
| SARS | Threshold (µW) | 1166.12 ± 20.23 | 1095.62 ± 16.29 | 982.72 ± 53.48 | 568.78 ± 15.51 | 681.47 ± 15.43 | 562.87 ± 10.14 | 620.19 ± 49.48 | 572.29 ± 49.26 | 616.89 ± 25.21 |
| | Efficiency (x $10^{-2}$ %) | 1.29 ± 0.13 | 0.84 ± 0.04 | 0.69 ± 0.02 | 15.09 ± 0.21 | 14.56 ± 0.31 | 14.32 ± 0.14 | 11.41 ± 0.19 | 15.52 ± 0.50 | 15.68 ± 0.41 |
| SRS | Threshold (µW) | 514.69 ± 57.83 | 605.51 ± 30.64 | 661.61 ± 30.50 | 339.20 ± 30.59 | 301.86 ± 8.72 | 348.97 ± 18.53 | 379.43 ± 25.09 | 276.79 ± 17.41 | 372.20 ± 9.14 |
| | Efficiency (%) | 3.51 ± 0.07 | 3.44 ± 0.13 | 3.18 ± 0.10 | 37.38 ± 1.13 | 37.86 ± 0.23 | 33.42 ± 0.25 | 28.90 ± 1.21 | 38.25 ± 0.70 | 32.50 ± 0.35 |

## References

1. E. D. Potter, J. L. Herek, S. Pedersen, Q. Liu, and A. H. Zewail, "Femtosecond laser control of a chemical reaction," Nature **355**, 66–68 (1992).
2. M. Motzkus, S. Pedersen, and A. Zewail, "Femtosecond real-time probing of reactions .19. Nonlinear (DFWM) techniques for probing transition states of uni- and bimolecular reactions," J. Phys. Chem. **100**, 5620–5633 (1996).
3. A. Assion, T. Baumert, M. Bergt, T. Brixner, B. Kiefer, V. Seyfried, M. Strehle, and G. Gerber, "Control of chemical reactions by feedback-optimized phase-shaped femtosecond laser pulses," Science **282**, 919–922 (1998).
4. M. Shim, B. Wilson, E. Marple, and M. Wach, "Study of fiber-optic probes for in vivo medical Raman spectroscopy," Appl. Spectrosc. **53**, 619–627 (1999).
5. E. Hanlon, R. Manoharan, T. Koo, K. Shafer, J. Motz, M. Fitzmaurice, J. Kramer, I. Itzkan, R. Dasari, and M. Feld, "Prospects for in vivo Raman spectroscopy," Phys. Med. Biol. **45**, R1–R59 (2000).
6. K. Kong, C. Kendall, N. Stone, and I. Notingher, "Raman spectroscopy for medical diagnostics - From in-vitro biofluid assays to in-vivo cancer detection," Adv. Drug Deliv. Rev. **89**, 121–134 (2015).
7. M. Islam, "Raman amplifiers for telecommunications," IEEE J. Sel. Top. Quantum Electron. **8**, 548–559 (2002).
8. M. Gonzalez-Herraez, S. Martin-Lopez, P. Corredera, M. Hernanz, and P. Horche, "Supercontinuum generation using a continuous-wave Raman fiber laser," Opt. Commun. **226**, 323–328 (2003).
9. J. Sharping, Y. Okawachi, and A. Gaeta, "Wide bandwidth slow light using a Raman fiber amplifier," Opt. Express **13**, 6092–6098 (2005).
10. T. J. Kippenberg, S. M. Spillane, D. K. Armani, and K. J. Vahala, "Ultralow-threshold microcavity Raman laser on a microelectronic chip," Opt. Lett. **29**, 1224–1226 (2004).
11. P. Latawiec, V. Venkataraman, M. J. Burek, B. J. M. Hausmann, I. Bulu, and M. Loncar, "On-chip diamond Raman laser," Optica **2**, 924–928 (2015).
12. X. Liu, C. Sun, B. Xiong, L. Wang, J. Wang, Y. Han, Z. Hao, H. Li, Y. Luo, J. Yan, T. Wei, Y. Zhang, and J. Wang, "Integrated continuous-wave aluminum nitride Raman laser," Optica **4**, 893–896 (2017).
13. G. Wang, M. Zhao, Y. Qin, Z. Yin, X. Jiang, and M. Xiao, "Demonstration of an ultra-low-threshold phonon laser with coupled microtoroid resonators in vacuum," Photonics Res. **5**, 73–76 (2017).
14. S. H. Huang, X. Jiang, B. Peng, C. Janisch, A. Cocking, Ş. K. Özdemir, Z. Liu, and L. Yang, "Surface-enhanced Raman scattering on dielectric microspheres with whispering gallery mode resonance," Photonics Res. **6**, 346–356 (2018).
15. C. Raman and K. Krishnan, "A new type of secondary radiation," Nature **121**, 501–502 (1928).
16. S. M. Spillane, T. J. Kippenberg, and K. J. Vahala, "Ultralow-threshold Raman laser using a spherical dielectric microcavity," Nature **415**, 621 (2002).
17. K. Georgiou, R. Jayaprakash, A. Askitopoulos, D. M. Coles, P. G. Lagoudakis, and D. G. Lidzey, "Generation of Anti-Stokes Fluorescence in a Strongly Coupled Organic Semiconductor Microcavity," ACS Photonics **5**, 4343–4351 (2018).
18. G. Eckhardt, D. Bortfeld, and M. Geller, "Stimulated emission of Stokes and anti-Stokes Raman lines from diamond, calcite, and alpha-sulfur single crystals," Appl. Phys. Lett. **3**, 137–138 (1963).
19. D. Leach, R. Chang, and W. Acker, "Stimulated anti-Stokes Raman scattering in microdroplets," Opt. Lett. **17**, 387–389 (1992).
20. D. Farnesi, F. Cosi, C. Trono, G. C. Righini, G. N. Conti, and S. Soria, "Stimulated anti-Stokes Raman scattering resonantly enhanced in silica microspheres," Opt. Lett. **39**, 5993–5996 (2014).
21. I. Itzkan and D. A. Leonard, "Observation of coherent anti−Stokes Raman scattering from liquid water," Appl. Phys. Lett. **26**, 106–108 (1975).
22. B. Hudson, W. Hetherington, S. Cramer, I. Chabay, and G. K. Klauminzer, "Resonance enhanced coherent anti-Stokes Raman scattering," Proc. Natl. Acad. Sci. **73**, 3798 (1976).
23. J. J. Barrett and R. F. Begley, "Low-power cw generation of coherent anti-Stokes Raman radiation in CH4 gas," Appl. Phys. Lett. **27**, 129–131 (1975).
24. K. Rittner, A. Hope, T. Muller-Wirts, and B. Wellegehausen, "Continuous anti-Stokes Raman lasers in a He-Ne laser discharge," IEEE J. Quantum Electron. **28**, 342–347 (1992).
25. Y.-Y. Cai, E. Sung, R. Zhang, L. J. Tauzin, J. G. Liu, B. Ostovar, Y. Zhang, W.-S. Chang, P. Nordlander, and S. Link, "Anti-Stokes Emission from Hot Carriers in Gold Nanorods," Nano Lett. **19**, 1067–1073 (2019).

26. N. Deka, A. J. Maker, and A. M. Armani, "Titanium-enhanced Raman microcavity laser," Opt. Lett. **39**, 1354–1357 (2014).
27. H. Choi and A. M. Armani, "High efficiency Raman lasers based on Zr-doped silica hybrid microcavities," ACS Photonics **3**, 2383–2388 (2016).
28. T. Kippenberg, S. Spillane, B. Min, and K. Vahala, "Theoretical and experimental study of stimulated and cascaded Raman scattering in ultrahigh-Q optical microcavities," IEEE J. Sel. Top. Quantum Electron. **10**, 1219–1228 (2004).
29. M. Oxborrow, "Traceable 2-D finite-element simulation of the whispering-gallery modes of axisymmetric electromagnetic resonators," IEEE Trans. Microw. Theory Tech. **55**, 1209–1218 (2007).
30. H. S. Choi, S. Ismail, and A. M. Armani, "Studying polymer thin films with hybrid optical microcavities," Opt. Lett. **36**, 2152–2154 (2011).
31. A. J. Maker, B. A. Rose, and A. M. Armani, "Tailoring the behavior of optical microcavities with high refractive index sol-gel coatings," Opt. Lett. **37**, 2844–2846 (2012).
32. Ph. Colomban and A. Slodczyk, "Raman intensity: An important tool in the study of nanomaterials and nanostructures," ACTA Phys. Pol. A **116**, 7–12 (2009).
33. D. Armani, T. Kippenberg, S. Spillane, and K. Vahala, "Ultra-high-Q toroid microcavity on a chip," Nature **421**, 925–928 (2003).
34. M. L. Gorodetsky, A. A. Savchenkov, and V. S. Ilchenko, "Ultimate Q of optical microsphere resonators," Opt. Lett. **21**, 453–455 (1996).
35. D. Hollenbeck and C. D. Cantrell, "Multiple-vibrational-mode model for fiber-optic Raman gain spectrum and response function," J. Opt. Soc. Am. B **19**, 2886–2892 (2002).
36. C. Evans, E. Potma, M. Puoris'haag, D. Cote, C. Lin, and X. Xie, "Chemical imaging of tissue in vivo with video-rate coherent anti-Stokes Raman scattering microscopy," Proc. Natl. Acad. Sci. U. S. A. **102**, 16807–16812 (2005).
37. X. Nan, E. O. Potma, and X. S. Xie, "Nonperturbative chemical imaging of organelle transport in living cells with coherent anti-stokes Raman scattering microscopy," Biophys. J. **91**, 728–735 (2006).
38. C. L. Evans, X. Xu, S. Kesari, X. S. Xie, S. T. C. Wong, and G. S. Young, "Chemically-selective imaging of brain structures with CARS microscopy," Opt. Express **15**, 12076–12087 (2007).
39. R. Riedinger, A. Wallucks, I. Marinković, C. Löschnauer, M. Aspelmeyer, S. Hong, and S. Gröblacher, "Remote quantum entanglement between two micromechanical oscillators," Nature **556**, 473–477 (2018).
40. T. T. Tran, B. Regan, E. A. Ekimov, Z. Mu, Y. Zhou, W. Gao, P. Narang, A. S. Solntsev, M. Toth, I. Aharonovich, and C. Bradac, "Anti-Stokes excitation of solid-state quantum emitters for nanoscale thermometry," Sci. Adv. **5**, eaav9180 (2019).